\documentclass[10pt, a4paper]{article}

\usepackage{tikz}

\usepackage{graphicx}

\usepackage{fancyhdr}

\usepackage{multicol}

\usepackage{multirow}

\usepackage{amsmath}

\usepackage{physics}

\usepackage{mhchem}

\usepackage{float}

\usepackage[colorlinks=true,citecolor=blue,pdftex]{hyperref}

\usepackage{siunitx}

\hypersetup{hidelinks,
backref=true,
pagebackref=true,
hyperindex=true,
breaklinks=true,
colorlinks=true,
urlcolor=blue,
bookmarks=true,
bookmarksopen=false,
linkcolor=blue,}

\usepackage{indentfirst}

\usepackage{graphicx}
\graphicspath{ {F:/Mirza/Uhamka/JO/Vol3No2/Images/} } 

\begin{document}\thispagestyle{empty}
\begin{multicols}{2}

\noindent
{\tiny ISSN: 2502-2318 (Online)\\*[-.4pc]
ISSN: 2443-2911 (Print)\\*[-.1pc]
Homepage: http://omega.uhamka.ac.id/}

\begin{flushright}
\begin{tikzpicture}
\draw (0,0) -- (1,0) -- (1,1) -- (0,1) -- cycle;
\draw (1,0) -- (4,0) -- (4,1) -- (1,1);
\draw (0.5,0.5) node{\LARGE{$\omega$}};
\draw (2.5,0.45) node{\LARGE{o m e g a}};
\end{tikzpicture}
\end{flushright}

\end{multicols}

\ \\

\begin{center}
Omega: Jurnal Fisika dan Pendidikan Fisika {\bf 5} (2), 1 - 2 (2019)\\
{\small ({\it Journal of Physics and Physics Education})}\\*[2pc]
{\Large Integration of Ocean Literacy into Physics Learning}\\*[1pc]
{\large Purwoko Haryadi Santoso$^{1,*}$}\\*[1pc]
{\small $^1${\it Department of Physics Education, Universitas Sulawesi Barat}}\\*[-.1pc]
{\small {\it Jl. Prof. Dr. Baharuddin Lopa SH, Talumung, Majene, Sulbar, 91412, Indonesia}}\\*[.1pc]
\end{center}
\rule{16cm}{0.1pt}\\*[-.5pc]

\noindent
{\bf Abstract}

\noindent
{\small Indonesia will be projected to be maritime axis in the world. Indonesian potential as a maritime country inspires Nawacita vision established by Joko Widodo in his cabinet. This ambitions must be supported by coastal communities. The province of West Sulawesi has a long coastal territory than other provinces in Indonesia. It should be exploited to promote the realization of Nawacita for the better Indonesia in the future. In education sector, 2013 curriculum introduces that teachers must implement scientific approach to their learning. Physics is one of the subjects that is still considered difficult by students. It will be more complex when it is taught by scientific approach. In other hand, physics is one of the most important subjects in preparing high school students for their job in science and technology. Contextual learning could be a solution in delivering physics to students. Without forgetting the 2013 curriculum, physics learning in the context of ocean literacy owned by West Sulawesi can lead students to gain meaningful experience. Based on these considerations, this article will discuss how to integrate ocean literacy into physics learning. }\\*[-.5pc]

\noindent
{\tiny \textcopyright \ 2019 The Authors. Published by Pendidikan Fisika UHAMKA}\\*[-.4pc]


\noindent
{\small DOI: \href{https://doi.org}{https://doi.org/...}}

\noindent
\rule{16cm}{0.1pt}

\noindent
{\tiny $^*$Corresponding author. E-mail address: purwokoharyadisantoso@unsulbar.ac.id}\\*[.5pc]

\begin{multicols}{2}
\section*{Introduction}

Contextual learning has been shown to increase student's interest in learning physics. This is brought about by the characteristic feature of contextual learning, which is to take everyday problems in studying physics. The teacher sees the potential possessed by the area where he teaches to be applied to the classroom. Through, examples that are easily found in real life will provide a more in-depth experience for students.

Based on data from the Central Statistics Agency, West Sulawesi province has considerable maritime potential. This shows that the province of West Sulawesi is a province that can support the ideals of the Indonesian state as a world maritime axis. This idea must be instilled in students as the next generation for the people of West Sulawesi. The hope is that through these students West Sulawesi province can take advantage of this maritime potential.

The potential of West Sulawesi must be used by the teacher in herding students in contextual physics learning. The teacher must sort out any maritime problems that can be studied in physics. Not all physics material can be integrated maritime insight.
Therefore, through this paper, the author will provide some problems in the context of maritime insight which according to the author can be integrated in several physical concepts such as vector analysis, straight motion, newton dynamics, rotational dynamics, rigid body equilibrium, and static fluid.

Students in physics learning usually do not get learning experience in the context of maritime. Most students cannot connect the skills and knowledge of physics with competencies related to maritime insight such as applications on radar, navigation, ship equilibrium, statics, dynamics, and other skills that students will need in exploiting West Sulawesi's regional potential. The lack of understanding of students in real life contexts that are close to physics results in the low interest and motivation of students having in studying physics. As a result, the learning outcomes of students become unsatisfactory.
Research in the field of teaching and learning has successfully investigated the difficulties of students in physics and mathematics in a few years. Educational experts suggest that learning methods on physics subjects be conveyed by linking them to the application of real life.

The benefits of studying physics in the context of life have long been known to education experts. In 1975, Carl R. Nave, Ph.D Physics students, and Brenda C. Nave, a professional nurse, collaborated in writing a book Physics for the Health Science. This paper is a university level textbook designed for nursing and medical students. In the introductory part of the book, the two authors state that "the writing of this book is driven by the desire that nursing and medical students must have a background in physics which will be applied in the need to carry out their profession later. This book is not just an introductory textbook on basic physics with its application in biology, but this book still discusses concepts found in basic physics books "\cite{1}. The couple of one of the authors, Rosemarie Pecota, also uses this book because she is a nursing student. He found that this book was very suitable and helpful for nursing students. However, when he changed his major to Chemistry, he felt that the Physics for the Health Sciences book was not enough to help students because it was too difficult to understand for general study programs such as Chemistry. Physics learning with a contextual model has a weakness that is naturally limited in its scope and can only be applied successfully to the fields being studied by students.

However, research shows that there is an increase in students' motivation when a contextual learning model is applied. Roth and Roychouldhury \cite{2} and Stinner \cite{3} show that "the context that is designed right can increase students' interest in learning physics" \cite{4}. Whitelegg \cite{5} found that making physics learning in a real life context can make participants not only learn abstract concepts of physics that do not understand their application in life, but students can understand physics as a way of thinking in real life. In addition, Hart \cite{6} found that "the context of life can involve and motivate students, explain the relevance and purpose of learning physics, and increase the level of mastery of students in learning". Other researchers write that "... physics must be practiced in a context that will enhance understanding through direct relevance to real life" \cite{7}.

Based on these considerations, the author will try to integrate maritime insight as the context of the physical problems examined in this article. The physics problems designed in this article will later be applied in physics learning in the classroom.

\section*{Vector Analysis in the Context of Ocean Literacy}

The concept of vector analysis can be conveyed to students by giving examples of relative motion that occurs when a ship is sailing in the ocean. The speed of sailing ships in the sea is greatly influenced by other styles such as the motion of the flow of sea water currents, the force of wind blowing, and the friction force between the water and the body of the ship.

\noindent
{}

\noindent
{\it Ocean Issues 1}

"A ship intends to move north at a speed of 40 km/h if there are no obstacles such as the force of wind blowing, the flow force of the sea water flow, and friction forces. However, reality says something else. Sea water flow leads 37$^o$ (measured towards north) to the northwest with a speed of 15 km/h. Meanwhile, the wind blows to the southeast as far as 53$^o$ (measured towards the north) by 8 km/h. In addition, there is friction between the body of the ship and water, the direction of which is always opposite to the motion of the ship by 10 km/h. What is the direction of movement of the ship?"

\noindent
{\it Solution : }

\begin{figure}[H]
\centering
\includegraphics[scale=.4]{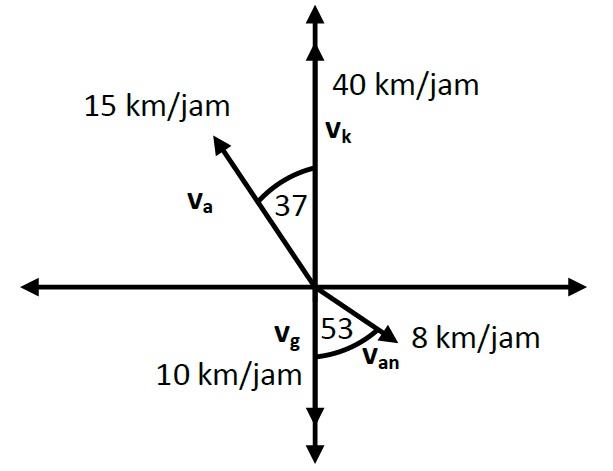}
\caption{\small Illustration of Ocean Issues 1}
\label{fig1}
\end{figure}

In solving this problem, of course students must have understood how to culture vectors as shown in Figure \ref{fig1}. Learners can apply the drawing method or analytical method in finding where the ship ends. The teacher can ask students to use both methods. Students should produce the same value even though the method of consulting it is different.

\begin{figure}[H]
\centering
\includegraphics[scale=1.5]{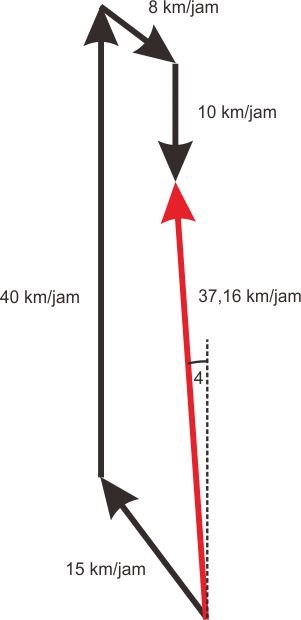}
\caption{\small Vector Resultant of Ocean Issues 1}
\label{fig2}
\end{figure}

First, the drawing method. Learners must have provided graph paper, bar, and bow. The method will be shown in Figure \ref{fig2}. Figure \ref{fig2} is analyzed using Corel DRAW X7 software. However, in implementing classroom learning, students must work on paper millimeters.

\begin{figure}[H]
\centering
\includegraphics[scale=.4]{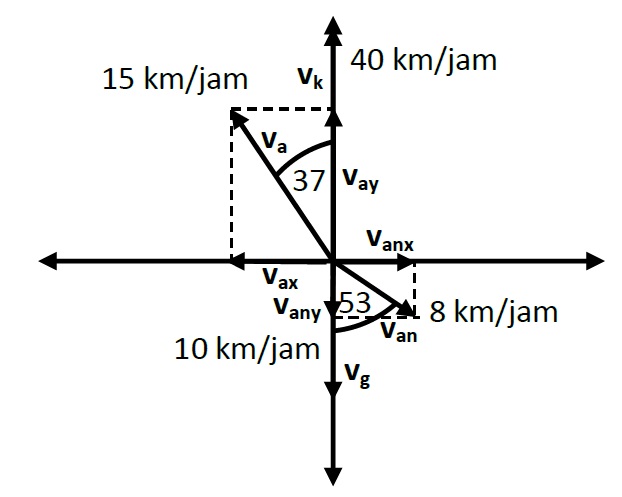}
\caption{\small Projection of vectors on the x-axis and y-axis}
\label{fig3}
\end{figure}

Second, analytical methods. Based on Figure \ref{fig2}, the resultant obtained is the speed of the ship is 32.16 km/h with the direction to the northwest with an angle of 4$^o$ towards the north. Now we will prove it by using an analytical method, namely by describing the existing forces into force components on the x-axis and y-axis. The results of the analysis are shown in Figure \ref{fig3}.

We obtain that,
\begin{equation}
\begin{split}
v_{ax} = v_a \sin 37 = (15)(0.6) = 9 \ \textrm{km/h}\\
v_{anx} = v_{an} \sin 53 = (8)(0.8) = 6.4 \ \textrm{km/h}
\end{split}
\end{equation}
\begin{equation}
\begin{split}
v_{ay} = v_a \cos 37 = (15)(0.8) = 12 \ \textrm{km/h}\\
v_{any} = v_{an} \cos 53 = (8)(0.6) = 4.8 \ \textrm{km/h}
\end{split}
\end{equation}

We result in the results of equations (1) and (2) on each axis, obtained
\begin{equation}
\Sigma{v_x}= v_{anx}-v_{ax}= 6.4-9 = -2.6 \ \textrm{km/h}
\end{equation}
\begin{equation}
\Sigma{v_y}=v_x+{v_ay}-{v_{any}}-{v_g}=37.2 \ \textrm{km/h}
\end{equation}

Based on equations (3) and (4), we get the resultant velocity of the ship.
\begin{equation}
\Sigma{R}=\sqrt{(-2.6)^2+(37.2)^2}=37.2 \ \textrm{km/h}
\end{equation}

The direction of the ship's final motion.
\begin{equation}
\tan\theta=\frac{\Sigma{v_x}}{\Sigma{v_y}}
\end{equation}

By entering equations (3) and (4) into equation (6), we get

\noindent \begin{center}
$$\tan\theta=\frac{37.2}{-2.6}=-14.3$$

$\theta=\arctan(-14.3)=94^o$ (terhadap x+)
\end{center}

\begin{equation}
\theta=4^o \textrm {(terhadap y+)}
\end{equation}

It is evident that what we find in results (5) and (7) matches what we are looking for through the graph method.

\section*{Linear Motion in the Context of Ocean Literacy}

\noindent
{\it Ocean Issues 2}

"Sea toll is an area/ island with several ships provided to collect goods collectively from several islands in areas that are difficult to reach. This idea will save ship fuel that will deliver goods to an area because it is sent collectively by one ship. Suppose there are 3 ships from 3 regions each, namely ship A, ship B and ship C. If ship A has a maximum speed of 100 km/h, ship B has a maximum speed of 80 km/h, and ship C has a speed of 120 km/h. The location of the sea toll is determined based on fairness in terms of the distance and capability of each vessel's engine to reach the sea toll port. So, the question is where is the sea highway from each port and what is the travel time of each ship to meet together on the sea highway?"

\noindent
{\it Solution : }

The third requirement to meet is that the three ships must travel the same time. Mathematically, we can write in equation (8)
\begin{equation}
t_A=t_B=t_C
\end{equation}

$$\frac{s_A}{100}=\frac{s_B}{80}=\frac{s_C}{120}$$

$$\frac{s_A}{100}\times\frac{12}{12}=\frac{s_B}{80}\times\frac{15}{15}=\frac{s_C}{120}\times\frac{10}{10}$$

$$12s_A=15s_B=10s_C$$

\noindent
So,
\begin{equation}
s_A:s_B:s_C=5:4:6
\end{equation}

Equation (9) shows the simplest comparison of the values of the three ships. For example, the port distances of ships A, B and C from sea tolls are 200 km, 160 km and 240 km, respectively.

The travel time of each ship is

$$t=\frac{s_A}{100}=\frac{s_B}{80}=\frac{s_C}{120}$$
\begin{equation}
\frac{200}{100}=\frac{160}{80}=\frac{240}{120}= 2 \ \textrm h
\end{equation}

\section*{Projectile Motion in the Context of Ocean Literacy}

\noindent
{\it Ocean Issues 3}

"A fisherman who is 1.7 m tall is fishing in the sea using a fishing rod that is 1.2 m long. He planned to put the bait 10 meters in front of the boat. What is the minimum speed that must be given to the end of the rod so that the hook reaches the target?"

\noindent
{\it Solution : }
\begin{figure}[H]
\centering
\includegraphics[scale=.295]{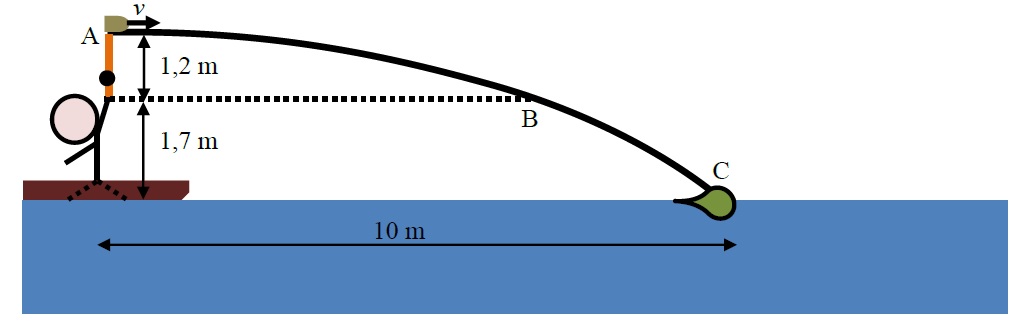}
\caption{\small Ocean Issues 3 Illustration}
\label{fig4}
\end{figure}

As shown in Figure \ref{fig4}, at point A

\begin{equation}
v_{ox}=v \ \textrm {dan} \     v_{oy}=0 
\end{equation}

Time to take A-B
\begin{equation}
y=v_{oy}t+\frac{1}{2}gt^2 
\end{equation}

$$1.2=0+\frac{1}{2}(10){t_{AB}}^2$$

\begin{equation}
t_{AB}=\sqrt{0.24}=0,49 \ \textrm s
\end{equation}

The component of the y-axis hook speed at point B is

\begin{equation}
v_y=v_{oy}+gt
\end{equation}

$$v_{yB}=0+(10)(0.49)=4,9 \ \textrm {m/s} $$

The time to take B-C, using equation (12), we get.

$$1.7=4.9t_{BC}+\frac{1}{2}(10){t_{BC}}^2$$
\begin{equation}
5{t_{BC}}^2+4,9t_{BC}-1.7=0
\end{equation}

Equation (14) is a quadratic equation that we can find a solution using the ABC formula

\begin{equation}
t_{BC1,2}=\frac{-b\pm\sqrt{b^2-4ac}}{2a}
\end{equation}

$$t_{BC1,2}=\frac{-4.9\pm\sqrt{(-4.9)^2-4(5)(-1.7)}}{10}$$

$$t_{BC1,2}=\frac{-4.9\pm\sqrt{58.01}}{10}$$

$$t_{BC1,2}=\frac{-4.9\pm{7.61}}{10}$$

\begin{equation}
t_{BC1}=\frac{-4.9+{7.61}}{10}=0,27 \ \textrm {s} 
\end{equation}

$$\textrm {or}$$

$$t_{BC2}=\frac{-4.9-{7.61}}{10}=-1.25 \ \textrm {s (impossible)}$$

So, by summing the results of equations (13) and (17), we get,

$$t_{AC}=t_{AB}+t_{BC}=0.76 \ \textrm {s}$$

The speed of the hook needed for a target within 10 meters is

$$v_{ox}=\frac{x}{t}=\frac{10}{0.76}=13.2 \ \textrm {m/s}$$

\section*{Newtonian Dynamics in the Context of Ocean Literacy}

\noindent
{\it Ocean Issues 4}

"If the ship is about to stop in the middle of the sea, the captain of the ship will instruct the sailors to lower the anchor so that the ship does not shift. One day, a ferry with a mass of 400 tons along with crew and ship goods was stopped at the Makassar Strait. At that time the sea water moves at an acceleration of 2 m/s$^2$. If it is known that the static friction coefficient between the iron anchor and the seabed is 0.8. What is the minimum anchor mass in order to hold the ship above the sea?"

\noindent
{\it Solution : }

As an equilibrium condition, the impulsive force caused by sea water must be the same as the static friction that the anchor has on the seabed. First, we first calculate the static friction force of the anchor used.
\begin{equation}
f=\mu N
\end{equation}
\begin{equation}
f=(0.8)(m)=0.8m
\end{equation}

Impulsive force that is driven by the flow of sea water

\begin{equation}
F=ma=4000\times2=800000 \ \textrm {N}
\end{equation}

\noindent
Required anchor mass,

$$\Sigma F=0$$
$$F-f=0$$
$$800000=0.8m$$
$$m=\frac{800000}{0.8}=1000 \ \textrm {tons}$$

\section*{Fluid Statics in the Context of Ocean Literacy}

\noindent
{\it Ocean Issues 5}

"A prism-shaped ship is an equilateral triangle. The ship's length is 100 m and the ship's face is an equilateral triangle having a side length of 20 m. If the ship can accommodate a maximum load of 1000 tons of cargo. How deep does the ship enter the sea water? ($\rho$ sea water = 1000 kg/m $^3$)"

\noindent
{\it Solution : }

At floating conditions,
\begin{equation}
F_A=w
\end{equation}

$$\rho_fV_{bf}g=mg$$
$$(1000)V_bf=1000000$$
$$V_bf=1000 \ \textrm {m$^3$}$$

\noindent
The area of the ship's face dipped into the sea.
\begin{equation}
L_{af}t=1000
\end{equation}
$$L_{af}(100)=1000$$
$$L_{af}=100 \ \textrm {m$^2$}$$

Consider the illustration of the ship's face in Figure \ref{fig5} below.

\begin{figure}[H]
\centering
\includegraphics[scale=.295]{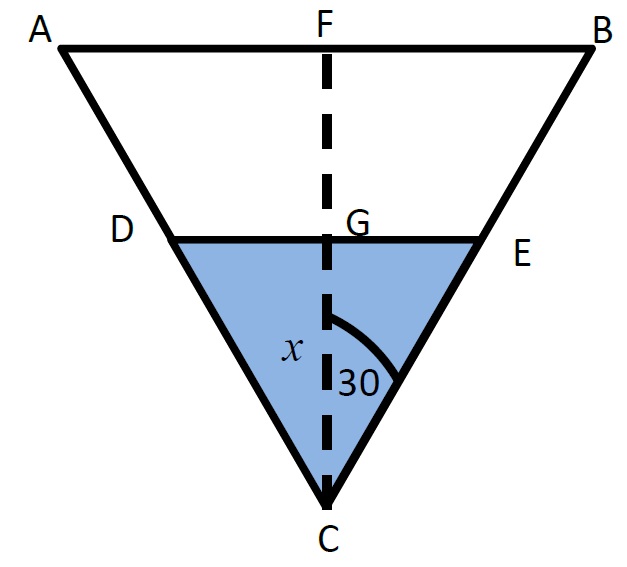}
\caption{\small Ship Advance Illustration}
\label{fig5}
\end{figure}

Note $\Delta$CGE in Figure \ref{fig5}. If we suppose GE is {\ it x}, then

$$\tan 30 = \frac{GE}{GC}$$

$$\frac{1}{\sqrt{3}}=\frac{x}{GC}$$

\begin{equation}
GC=x\sqrt{3}
\end{equation}

The depth of the ship that dipped into the water

$$\textrm {Area of} \ \Delta \textrm{CDE}=\frac{GE\times GC}{2}$$

$$10=\frac{x\times x\sqrt{3}}{2}$$

$$10=\frac{x^2\sqrt{3}}{2}$$

\begin{equation}
x=\sqrt{11.55}=3.39 \ \textrm{m}
\end{equation}

\section*{Conclusion}

Bringing the ocean literacy in physics learning is one step for physics educators to support the vision of exploiting the potential of West Sulawesi. Examples of problems that have been presented can be tested in a direct physics learning environment. This is done to analyze whether this idea can improve learning outcomes and motivation of students towards physics.

\end{multicols}

\end{document}